\documentclass{article}
\usepackage[utf8]{inputenc}
\usepackage[a4paper]{geometry}
\usepackage{amsmath,amssymb}
\usepackage{cite}
\usepackage{bm}
\usepackage{graphicx}
\usepackage{epsfig} 
\usepackage{comment}
\usepackage{color}
\usepackage{float}
\usepackage{bm}
\usepackage{caption}
\usepackage{subcaption}
\usepackage{comment}
\usepackage{authblk}

\usepackage[hidelinks]{hyperref}

 \allowdisplaybreaks
\usepackage{mathtools}
\DeclareMathOperator{\Tr}{Tr}
\newcommand{\av}[1]{\left< #1 \right>}
\DeclarePairedDelimiter\bra{\langle}{\rvert}
\DeclarePairedDelimiter\ket{\lvert}{\rangle}
\DeclarePairedDelimiterX\braket[2]{\langle}{\rangle}{#1 \delimsize\vert #2}
\DeclareMathOperator{\sech}{sech}

\title{Thermodynamics of one and two-qubit quantum refrigerators interacting with squeezed baths: a comparative study}
\author[1]{Ashutosh Kumar \thanks{ashutoshkumarr06@gmail.com}}
\author[1]{Sourabh Lahiri \thanks{sourabhlahiri@gmail.com}}

\affil[1]{Birla Institute of Technology Mesra, Ranchi, Jharkhand 835215, India}

\date{}

\begin{document}

\maketitle
\begin{abstract}
    We investigate the nonequilibrium refrigeration of one and two-qubit systems in a squeezed thermal bath. We characterize the performance of one and two-qubit refrigerators in the presence of squeezed heat baths, in terms of their coefficients of performance, cooling rates, and figures of merit.
   Our results show that the performance of the refrigerators is strongly influenced by the squeezing parameter and the number of qubits. The performance of the two-qubit refrigerator is found to be  better than that of the one-qubit refrigerator under the same operating conditions. Our findings suggest that a squeezed thermal bath can be a promising resource for the design of efficient quantum refrigerators in the non-equilibrium regime.
\end{abstract}

\section{Introduction}

Refrigerators are thermal devices that are useful for cooling a system. 
Over the last two decades, there has been a lot of study towards extending these macroscopic machines to microscopic scales \cite{blickle2012realization,schmiedl2007efficiency,kumari2020stochastic,kumari2021microscopic,long2015performance}. They are likely to have a significant effect in energy harvesting at nanoscales and are predicted to have the potential of revolutionizing the pharmaceutical industry \cite{bhat2014nanobots,manjunath2014promising,freitas2006pharmacytes}. Tiny refrigerators have been shown to be useful in cooling AFM cantilevers \cite{liang2000thermal} and in intramolecular cooling \cite{briegel2008entanglement}.

 Typically, a Brownian particle trapped in a confining potential is used in classical stochastic thermal machines \cite{rana2014single,schmiedl2007efficiency,blickle2012realization}. A lot of interest has been directed towards quantum heat engines, which have been shown to offer several advantages over their classical counterparts \cite{linden2010small,brunner2014entanglement,rossnagel2014nanoscale,abah2016optimal}. 
 Pioneering work on quantum heat engine (QHE) was done in \cite{scovil1959three} that dealt with maser heat engines.  In \cite{alicki1979quantum,kosloff1984quantum,zhang2020optimization,wang2013efficiency}, the authors proposed models of a QHE within an open quantum system framework.
 Quantum thermodynamics \cite{vinjanampathy2016quantum} has become a standard framework for studying these machines.
 
  A comparative study of the performance of one and two-qubit quantum engines in the presence of squeezed heat baths was done in \cite{lahiri2023thermodynamics}, demonstrating that the two-qubit engines typically yield higher power. Additionally, by adjusting the squeezing parameters, the machine can be made to operate in either engine or in refrigerator mode.
  
  The coefficient of performance (COP) is a measure of the quality of results delivered by a refrigerator or a heat pump. A higher COP indicates a more efficient system. 
  An ideal candidate for optimizing the performance  of a refrigerator is its figure of merit $\chi$, which was first studied in \cite{yan1990class}. It provides equal weightage to the COP of the refrigerator and its cooling rate ($CR$) and is an important parameter to determine how useful the device is.
   In \cite{singh2020optimal}, the authors study the optimization of $\chi$ and $CR$ for a three-level atomic system.

 Here, we study the one and two-qubit refrigerator in a nonequilibrium setup (finite cycle time) in the presence of squeezed thermal baths \cite{ficek2002entangled,banerjee2010dynamics}. This entails the use of the quantum master equation \cite{subhashish2019open,breuer2002theory}. The refrigerator protocol is obtained by time-reversing the Otto engine protocol of \cite{lahiri2023thermodynamics}, so that the clear distinctions between the steps where work is done and those where heat is exchanged are retained. The functional dependences of the thermodynamic observables of the refrigerator on various parameters are studied, first for the one-qubit and then for the two-qubit system. We compare the qualitative and quantitative natures of the observables and study their similarities and differences. Generally, we find enhanced performances when two qubits constitute the working system - a result that was also observed when the engine protocol was applied to the system \cite{lahiri2023thermodynamics}, even though a different set of thermodynamic observables were studied there.
 
The paper is organized as follows: In sec. \ref{sec:model}, we discuss the model of one-qubit and two-qubit refrigerators. In Sec. \ref{sec:thermodynamics}, we define the thermodynamic observables. In sec. \ref{sec:OQR}, we discuss the result of one-qubit refrigerators. Sec. \ref{sec:TQR} provides the corresponding study for two-qubit refrigerators. Finally, in Sec. \ref{sec:conclusion}, the key conclusions of this work are summarised and discussed.

\section{Model}
\label{sec:model}

In this section, we describe the models used to study quantum refrigerators using a single qubit or two qubits as its working system.
The refrigerator cycle considered is the reverse of the protocol used in \cite{lahiri2023thermodynamics} for a quantum heat engine, comprised of two thermodynamic processes - adiabatic and isochoric. Fig. \ref{fig:SchematicOtto} schematically shows such a cycle, given by the  process  A$\rightarrow$B$\rightarrow$C$\rightarrow$D$\rightarrow$A, labeled as strokes 1, 2, 3, and 4, respectively.

During the adiabatic compression step (A$\to$B), the energy gap between the two levels (for the two-qubit system, it is chosen as the mean energy gap of the two qubits) changes from  $\omega_h$ to $\omega_c$. The next step B$\to$C is the isochoric one, where the system evolves in contact with the cold bath maintained at temperature $T_c$. The third step C$\to$D consists of adiabatic expansion where  $\omega_c$ is changed to $\omega_h$, while the final step D$\to$A is another isochoric one where the system evolves in contact with the hotter bath at temperature $T_h$. As is clear from the description, the evolution during the adiabatic steps are unitary, while those in the isochoric steps are non-unitary. In addition, the heat baths are subjected to squeezing, which is described by means of the squeezing parameters $r$ and $\phi$ (to be elaborated below). We find the latter parameter to have no effects on the engine's output parameters. Further, the two heat baths are in general subject to different squeezing parameters: $r=r_h$ for the hotter bath, and $r=r_c$ for the colder bath. The effective temperature of the heat baths under the effect of squeezing can be shown to be given by $T^{\rm eff}_{h/c} = T_{h/c}\cosh(2r_{h/c})$, as has been shown in \cite{manzano2018squeezed} ($T_{h/c}$ is the actual thermal temperature of the hot/cold bath).

 The evaluation of the mean energy of the one-qubit and two-qubit systems at the four corners of the cycle is necessary for the examination of the performance of the refrigerator. The average energy at any state $\alpha$ is $\av{E_\alpha}\equiv \Tr[\rho_\alpha H_\alpha]$, $\rho_\alpha$ being the reduced density matrix for the system and $H_\alpha$ being the Hamiltonian operator, with $\alpha=$A, B, C or D. Further details are provided in Sec. \ref{sec:thermodynamics}.

\subsection{Model for a one-qubit refrigerator (OQR)}

\begin{figure}[!ht]
	\begin{center}
	\includegraphics[width=1.0\textwidth]{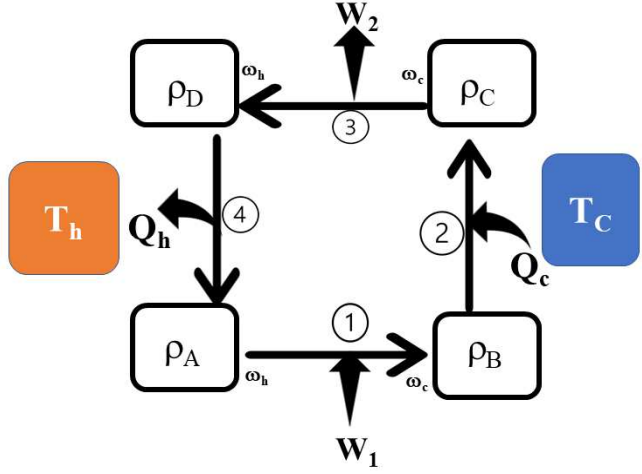}
    	\caption{An illustration of the one-qubit quantum Otto refrigerator cycle.}
		\label{fig:SchematicOtto}
	\setlength{\belowcaptionskip}{1pt}
	\end{center}
\end{figure}

The Hamiltonian of the one-qubit system, in the $z$-basis, is given by
\begin{align}
    H(\omega(t)) = \frac{1}{2}\omega(t)\sigma_z.
\end{align}
 Here, $\sigma_z$ is the third Pauli matrix, and $\omega(t)$ is the time-dependent energy gap between the two energy levels.
We now consider looking at the development of the density operator, which is represented by the von Neumann equations during the unitary steps A$\to$B and C$\to$D, and by the Lindblad equations during the dissipative steps B$\to$C and D$\to$A (see Fig. \ref{fig:SchematicOtto}). The density operator is a $2\times 2$ matrix, which will be represented in the energy basis $\{|g\rangle,|e\rangle\}$ that correspond to the ground and the excited states, respectively.

\paragraph{\bf Stroke 1, A $\to$ B (Adiabatic Compression):}  The frequency of the Hamiltonian is linearly modulated with time from $\omega{_h}$ to $\omega{_c}<\omega_h$ linearly in time:
 \begin{align}
\omega(t) &=\omega_h(1-t/\tau) +\omega_ct/\tau.
\label{eq:H_AB}
\end{align}
The von-Neumann equation is used to evolve the density matrix $\rho(t)$.
 \begin{align}
\frac{\partial \rho(t)}{\partial t} &=-\frac{i}{\hbar}[H(\omega(t)),\rho(t)],
\label{eq:VonNeumann}
\end{align}
where
\begin{align}
\sigma_z = \begin{pmatrix}1 & 0 \\
                0 & -1
            \end{pmatrix}.\\ 
\label{eq:H_BA}
\end{align}

The work done on the system in this process will be given by $W = \langle E_B\rangle - \langle E_A\rangle$.
For convenience, in this article, we simplify the equations by setting Planck's constant and Boltzmann's constant to one (i.e., $\hbar=1$, $k_B=1$).
\paragraph{Stroke 2, B $\to$ C (cold isochore):} This is a non-unitary process (since the system is in contact with the cold thermal bath), and the final Hamiltonian at C is given by $H(\omega_c)$ =$(\omega_c/2)\sigma_z$. Heat is absorbed from the cold bath by the system. The quantum master equation results in the evolution of the system's density matrix $\rho(t)$, which is shown below
\begin{equation}
	\begin{split}
		\frac{\partial \rho(t)}{\partial t} =-\frac{i}{\hbar}[H(\omega_c),\rho(t)]+\sum_{j=1,2}\left[2R_j\rho(t) R^\dagger_j - \left\{R^\dagger_j R_j,\rho(t)\right\} \right],
	\end{split}
	\label{eq:master}
\end{equation}
where $R_1=\left[\sqrt{\gamma_0(n_{th}+1)/2}\right]R$, $R_2=\left[\sqrt{\gamma_0 n_{th}/2}\right]R^\dagger$, and $R=\cosh(r)\sigma_{-}+\exp(i\phi)\sigma_{+}\sinh(r)$, with $r$ and $\phi$ being the squeezing parameters. The parameter $\gamma_0$ is the rate of spontaneous emission.  $n_{th}=1/(e^{\beta_c\omega_c}-1)$ gives the average number of photons emitted with frequency $\omega_c$ and inverse temperature $\beta_c$. The curly brackets $\{A_1,A_2\}$ denote anti-commutation of operators $A_1$ and $A_2$.

\paragraph{Stroke 3, C $\rightarrow$ D (Adiabatic expansion):} The system is now decoupled from the cold bath. The energy level spacing is changed linearly from $\omega{_c}$ to $\omega{_h}$ according to the protocol  
\begin{align}
    \omega(t)=\omega_h(t/\tau-2)+\omega_c(3-t/\tau).
    \label{eq:Omega_DC}
\end{align}
The density matrix $\rho(t)$ is again evolved using the von-Neumann equation (see Eq. \eqref{eq:VonNeumann}).

\paragraph{Stroke 4, D $\to$ A (Hot isochore):}The system is now coupled to the hot thermal bath. The Hamiltonian in this stroke is kept fixed at $H(\omega_h)=(\omega_h/2)\sigma_z$, with the energy gap held constant at $\omega_h$. Again, the evolution of the state $\rho(t)$ is through the quantum master equation (see Eq. \eqref{eq:master}). In this step, heat is released into the hot bath.

If the process is carried out quasistatically, the density operators at the states A, B, C, and D would correspond to the Boltzmann distribution, and the energy averages are easy to calculate in the absence of any squeezing:
\begin{align}
   \langle E_A\rangle &= \frac{-\omega_h}{2}\tanh\left(\frac{\beta_h\omega_h}{2}\right);\nonumber\\
    \langle E_B\rangle &= \frac{-\omega_c}{2}\tanh\left(\frac{\beta_h\omega_h}{2}\right); \nonumber\\
       \langle E_C\rangle &= \frac{-\omega_c}{2}\tanh\left(\frac{\beta_c\omega_c}{2}\right); \nonumber\\
    \langle E_D\rangle &= \frac{-\omega_h}{2}\tanh\left(\frac{\beta_h\omega_h}{2}\right).
    \label{eq:AverageEnergies}
\end{align}
However, throughout this work, we will generally be dealing with a nonequilibrium engine, and the average energy at these states will be different from the equilibrium ones.
   

\subsection{\label{subsec:two-qubit}Model for a two-qubit refrigerator (TQR)}

We now consider a refrigerator that uses a two-qubit system as the working medium that interacts with a squeezed thermal bath. An additional parameter emerges now, that would affect the performance of the refrigerator in this case. It is the distance $r_{12}$ between qubit 1 and qubit 2. When the qubits are close enough (see below), they interact collectively with the bath and are thus said to be in the collective regime.  For much larger distances between them, they are said to be in an independent regime.

Let the energy level spacings of qubits 1 and 2 be given by $\omega_1(t)$ and $\omega_2(t)$, respectively.  The evolution of the density matrix of the two-qubit system is given by \cite{subhashish2019open,ficek2002entangled,lahiri2023thermodynamics}
\begin{align}
   \frac{\partial \rho(t)}{\partial t} &=- \frac{i}{\hbar}[\tilde{H},\rho(t)]-L_1(\rho)-L_2(\rho)+L_3(\rho)+L_3(\rho),\nonumber\\
	\label{eq:two_master}
   \end{align}
  where the operator $\tilde H$ is as defined in Eq. \eqref{eq:two_hamiltonian}, and
 \begin{align*}
 L_1(\rho) &= \frac{1}{2}\Gamma_{12}\sum_{i,j=1}^{2}(1+N_{th}[\cosh^2(r)+\sinh^2(r)]+\sinh^2(r))(\rho S_i^+S_j^-+S_i^+S_j^-\rho-2S_j^-\rho S_i^+);\nonumber\\
 L_2(\rho) &= \frac{1}{2}\Gamma_{12}\sum_{i,j=1}^{2}(N_{th}[\cosh^2(r)+\sinh^2(r)]+\sinh^2(r))(\rho S_i^-S_j^+ +S_i^-S_j^+\rho -2S_j^+\rho S_i^-);\nonumber\\
 L_3(\rho) &= \frac{1}{2}\Gamma_{12}\sum_{i,j=1}^{2}(-\frac{1}{2}\sinh(2r)\exp{(i\phi)})(2N_{th}+1)(\rho S_i^+S_j^+ +S_i^+S_j^+\rho -2S_j^+\rho S_i^+);\nonumber\\
 L_4(\rho) &= \frac{1}{2}\Gamma_{12}\sum_{i,j=1}^{2}(-\frac{1}{2}\sinh(2r)\exp{(-i\phi)})(2N_{th}+1)\}(\rho S_i^-S_j^- +S_i^-S_j^-\rho -2S_j^-\rho S_i^-),\nonumber\\
   \end{align*}
   \begin{align}
       N_{th} &=\frac{1}{\exp(\hbar\omega_0/k_BT)-1}; \nonumber\\
     \omega_0 &=\frac{\omega_1+\omega_2}{2}.
   \end{align}

Here,  $\omega_{1h}$, $\omega_{2h}$, $\omega_{1c}$, and $\omega_{2c}$ correspond to the different level spacings of the independent one-qubit systems (just as in the case of the OQR), except that an extra index (1 and 2) indicate whether the state is of qubit 1 or qubit 2. We further define: $\omega_{0h}=(\omega_{1h}+\omega_{2h})/2$ and $\omega_{0c}=(\omega_{1c}+\omega_{2c})/2$. The time-dependence of the Hamiltonian comes through the parameter $\omega_{0}(t)$ that replaces $\omega(t)$ in Eq. \eqref{eq:Omega_DC}, where  $\omega_{0h}$ and $\omega_{0c}$ are substituted for $\omega_h$ and $\omega_c$, respectively.
   $r$ and $\phi$ are the squeezing parameters and $\Gamma_{12}$ is the collective spontaneous emission rate given in terms of individual spontaneous emission rates $\Gamma_1$ and $\Gamma_2$:
 \begin{align}
   \Gamma_{12}=\Gamma_{21} = \sqrt{\Gamma_1 \Gamma_2}~F(k_0 r_{12}), 
   \label{eq:Gamma}
  \end{align}
   where 
   \begin{align}
    \Gamma_i &=\frac{\omega_i^3\mu_i^2}{3\pi\epsilon\hbar c^3}, \hspace{0.5cm}(i=1,2), \hspace{0.5cm} r_{12}=|\bm{r}_{12}|=r_{21}, \hspace{0.5cm} k_0 = \omega_0/c, \nonumber\\
    \\
     F(k_0 r_{12}) &=\frac{3}{2}\left[\{1-(\bm{\hat{\mu}.\hat{r}}_{12})^2\}\frac{\sin{(k_0r_{12})}}{(k_0r_{12})}\right.\hspace{1cm}\nonumber\\
     &\left.+\{1-3(\bm{\hat{\mu}.\hat{r}}_{12})^2\}\times\left\{\frac{\cos{(k_0 r_{12})}}{(k_0 r_{12})}^2-\frac{\sin{(k_0 r_{12})}}{(k_0 r_{12})^3}\right\}\right].
    \end{align}
    Here, $\bm{r_{12}}\equiv (\bm{r}_2 - \bm{r}_1)\omega_0/c$ is the normalized displacement vector of the second spin with respect to the first. The dipole moments of atomic transition are given by $\bm{\mu}_1$ and $\bm{\mu}_2$. The hats represent unit vectors. 
  For identical qubits that are considered in our case, $\Gamma_1 = \Gamma_2 = \Gamma$ and $\mu_1 = \mu_2 = \mu$. 
 The dynamics of the system in contact with the cold bath follow similar equations. 
 The Hamiltonian $\Tilde{H}$ appearing in the Eq. \eqref{eq:two_master} is given by 
 \begin{equation}
\label{eq:two_hamiltonian}
  \tilde{H} = \hbar (\omega_1 S^{\rm z}_1 + \omega_2 S^{\rm^ z}_2)+
  \hbar \Omega_{12}(S^{\rm +}_1 S^{\rm -}_2 + S^{\rm +}_2 S^{\rm -}_1),\\
 \end{equation}
where 
  \begin{align}
S^{\rm z}_1  &= \frac{1}{2}(\ket{e_1}\bra{e_1}-\ket{g_1}\bra{g_1}); \hspace{1cm}
S^{\rm z}_2  = \frac{1}{2}(\ket{e_2}\bra{e_2}-\ket{g_2}\bra{g_2})\nonumber\\
S^{\rm +}_1  &= \ket{e_1}\bra{g_1},\hspace{0.2cm}  S^{\rm +}_2 = \ket{e_2}\bra{g_2};
\hspace{1cm} S^{\rm -}_1  = \ket{g_1}\bra{e_1},\hspace{0.2cm}  S^{\rm -}_2 = \ket{g_2}\bra{e_2}.\nonumber\\  
\Omega_{12}&=\Omega_{21}=\frac{3}{4}\Gamma\left[-\{1-(\bm{\hat{\mu}.\hat{r}}_{12})^2\} \frac{\cos{(k_0 r_{12})}}{(k_0 r_{12})}\right.\nonumber\\ 
  &\hspace{2cm}\left.+\{1-3(\bm{\hat{\mu}.\hat{r}}_{12})^2\}\times\left\{\frac{\sin{(k_0 r_{12})}}{(k_0 r_{12})}^2+\frac{\cos{(k_0 r_{12})}}{(k_0 r_{12})^3}\right\}\right].
  \end{align}

The convenient basis  that diagonalizes the $\tilde H$, also called the dressed state basis, is given by:
\begin{align}
\label{eq:dressed_basis}
      \ket{g} &= \ket{g_1}\ket{g_2},\nonumber\\
      \ket{s} &= \frac{1}{\sqrt{2}}(\ket{e_1}\ket{g_2} + \ket{g_1}\ket{e_2}),\nonumber\\
      \ket{a} &=\frac{1}{\sqrt{2}}(\ket{e_1}\ket{g_2} - \ket{g_1}\ket{e_2}),\nonumber\\
      \ket{e} &= \ket{e_1}\ket{e_2}.\nonumber\\
\end{align}
  The eigenvalues are respectively given by $E_g = -\hbar\omega_0$, $E_s = \hbar\Omega_{12}$, $E_a = -\hbar\Omega_{12}$ and $E_e = \hbar\omega_0$. The differential equations followed by each element of the density matrix are provided in \cite{banerjee2010dynamics,lahiri2023thermodynamics}. We simulate these equations to obtain the evolution of the density matrix.

\section{Important thermodynamic quantities} \label{sec:thermodynamics}
\begin{enumerate}
    \item {\bf Work:} 
    In the analysis of any heat engine/refrigerator, the output work is of fundamental importance. The unitary processes ($A \to B$ and $C \to D$) constitute the steps where work is done on/by the system. The works retrieved from the system in these two processes are given by
     \begin{align}
        \av{W_{AB}} &= \av{E_B}-\av{E_A}; \hspace{0.5cm}
       \av{W_{CD}} = \av{E_D} - \av{E_C}.
 \end{align}
\item {\bf Heat:} 
The heat absorbed by the system in the non-unitary steps is defined as (a negative value would indicated heat dissipated):
   \begin{align}
       \av{Q_{BC}} &= \av{E_C}-\av{E_B}; \hspace{1cm}          
       \av{Q_{DA}} = \av{E_A}-\av{E_D}.                
   \end{align}

  \item{\bf Coefficient of Performance:} 
  A refrigerator's \emph{coefficient of performance ($\zeta$)} is obtained by dividing the heat extracted from the cold reservoir by the total work:
       \begin{align}
          \zeta = \frac{\av{Q_c}}{\av{W}} = \frac{\omega_c}{\omega_h-\omega_c},
          \label{eq:OQR_cop}
       \end{align}
       whose derivation has been derived in appendix \ref{app:cop}.

\item{\bf Cooling rate:} 
It is defined as the heat absorbed from the cold bath divided by the cycle time:
       \begin{align}
            CR &=\frac{\av{Q_c}}{\tau_{\rm cycle}}
            \label{eq:cr_1}
        \end{align}  
        We define the maximum value of $CR$ with respect to  $\omega_c$ as the maximum cooling rate, $CR_{max}$ \cite{singh2020optimal}.

\item{\bf Figure of merit $(\chi)$:} 
The figure of merit $\chi$ is defined as the product of heat absorbed from the cold reservoir $\langle Q_c\rangle$ and the coefficient of performance ($\zeta$) divided by the cycle time \cite{yan1990class} :
\begin{align}
    \chi &=\frac{ \langle Q_c\rangle  {\zeta}}{\tau_{\rm cycle}}\nonumber\\
         &= CR \times \zeta.
\end{align}
%
\end{enumerate}

  \subsection{Quasistatically driven one-qubit refrigerator in absence of squeezing}
           We now derive a few analytical expressions in the absence of squeezing of either the hot or the cold baths ($r_h=r_c=0$), which we use as benchmarks for our simulations.
           If we use the expressions for equilibrium steady state in the absence of squeezing, namely Eq. \eqref{eq:AverageEnergies}, then the $CR$ is given by
        \begin{align}{\label{eq:Ana_cr_1q}}
            CR &=\frac{-\frac{1}{2}\omega_c\tanh\left(\frac{\beta_c\omega_c}{2}\right)+\frac{1}{2}\omega_c\tanh\left(\frac{\beta_h\omega_h}{2}\right)}{\tau_{\rm cycle}},
        \end{align}
        which leads to the maximum cooling rate $CR_{\rm max}$ by equating $\frac{d(CR)}{d\omega_c}$ to 0:
    \begin{align}
      CR_{\rm max} &=\frac{\frac{1}{4}\beta_c\omega^2_c\sech{\left(\frac{\beta_c\omega_c}{2}\right)^2}}{\tau_{\rm cycle}}.
      \label{eq:omega_max}      
\end{align}
    The expression for the figure of merit is given by
     \begin{align}
      \chi &=\frac{\omega_c\left[-\frac{1}{2}\left(\omega_c\tanh\left(\frac{\beta_c\omega_c}{2}\right)+\omega_c\tanh\left(\frac{\beta_h\omega_h}{2}\right)\right)\right]}{(\omega_h-\omega_c) \tau_{\rm cycle}}.
      \label{eq:chi} 
\end{align}
$\chi$ at the maximum cooling rate (MCR) can be readily computed. The following expression is obtained:
\begin{align}
      \chi_{\rm MCR} &=\frac{\omega_c\left[\frac{1}{4}\beta_c\omega^2_c\sech{\left(\frac{\beta_c\omega_c}{2}\right)^2}\right]}{(\omega_h-\omega_c) \tau_{\rm cycle}}.
      \label{eq:chi_max} 
\end{align}

\section{\label{sec:OQR}Results and Discussions for one-qubit refrigerator}

We now study the thermodynamics of the OQR and later use the results for comparison when we study the TQR in the next section. The squeezing parameter $\phi$ has no effect on the output, so we set them to zero throughout our analysis. Therefore, a reference to the squeezing parameter would solely refer to the parameter $r$ (with subscripts $h$ and $c$, in order to indicate the bath in whose contact the system is evolving) in Eq. \eqref{eq:master}.

\begin{figure}[!ht]
 \centering
 \begin{subfigure}{0.48\textwidth}
     \centering
     \includegraphics[width=\textwidth]{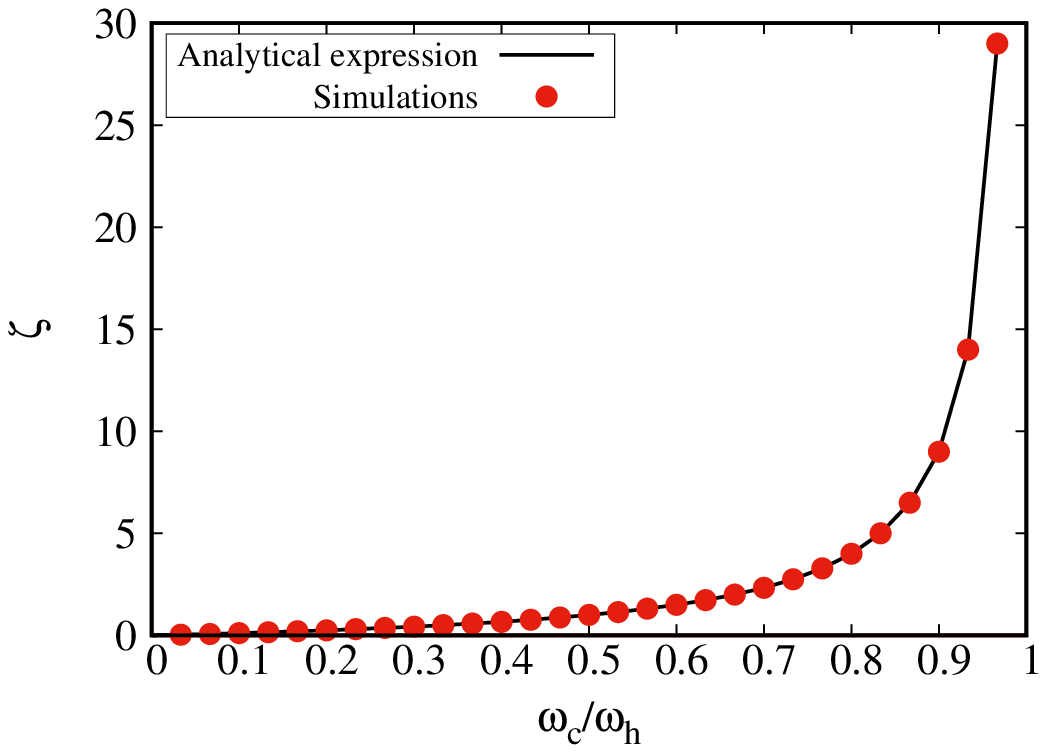}
     \caption{}
 \end{subfigure}
 	 \begin{subfigure}{0.48\textwidth}
     \centering
     \includegraphics[width=\textwidth]{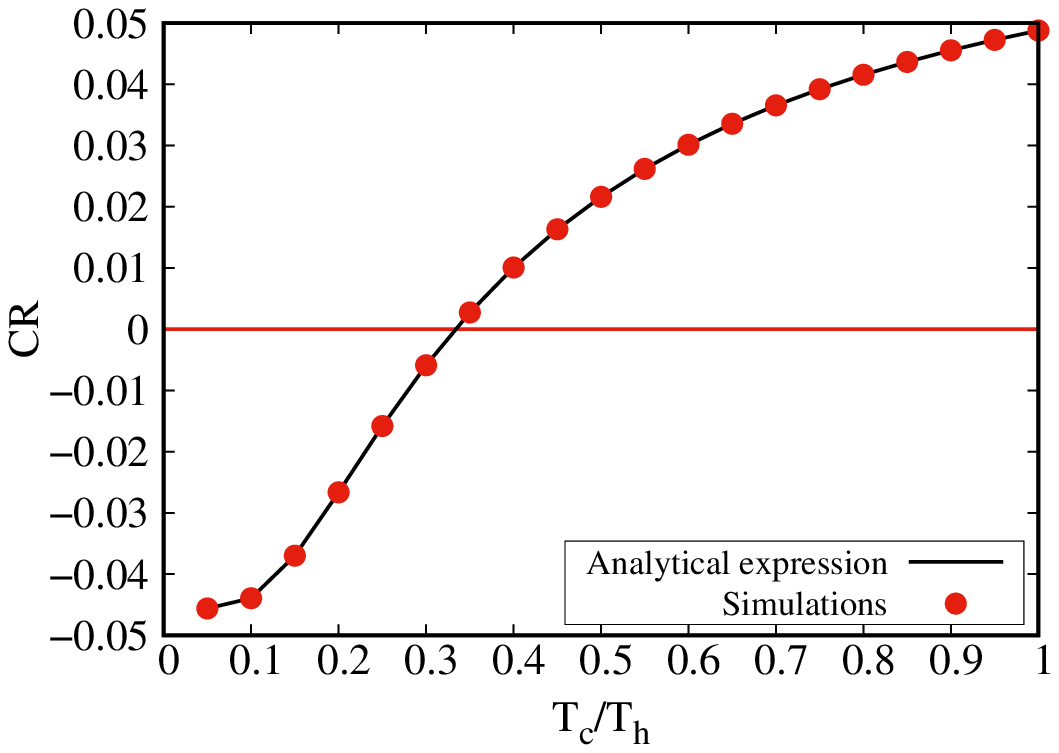}
     \caption{}
 \end{subfigure}
  \caption{ (a) The behavior of $\zeta$ as a function of $\omega_c/ \omega_h$, obtained from simulations (denoted by symbols) as well as analytics (solid line) in the absence of squeezing. (b) Comparison between analytical expression and simulations of $CR$ as a function of $T_c/T_h$ in the quasistatic limit  at $r_h = r_c = 0 $.  Other parameters are : $\omega_h=30$, $\omega_c=10$, $\gamma_0=1$, $\tau=10$.}
  \label{fig:benchmarking}
  \end{figure}

\paragraph{Comparison with analytics:}
 In order to benchmark our codes, we compare the analytical expressions obtained from Eqs. \eqref{eq:OQR_cop} and \eqref{eq:Ana_cr_1q} with the results obtained from our simulations in Figs. \ref{fig:benchmarking}(a) and (b), respectively. The symbols indicate results obtained from our simulations, while the solid line is the result obtained from analytics. Fig. \ref{fig:benchmarking}(a) shows an excellent agreement between the analytical (see Eq. \eqref{eq:OQR_cop}) and simulated values of the coefficient of performance $\zeta$. Similar degree of agreement is observed between the analytical (see Eq. \eqref{eq:Ana_cr_1q}) and simulated values of  $CR$ in Fig. \ref{fig:benchmarking}(b), in absence of squeezing ($r_h=0$, $r_c= 0$) and in the limit of quasistatic driving. This establishes the accuracy of our simulations, thereby benchmarking our code.
  
\begin{figure}[!ht]
 \centering
 	\begin{subfigure}{0.48\textwidth}
	\includegraphics[width=\textwidth]{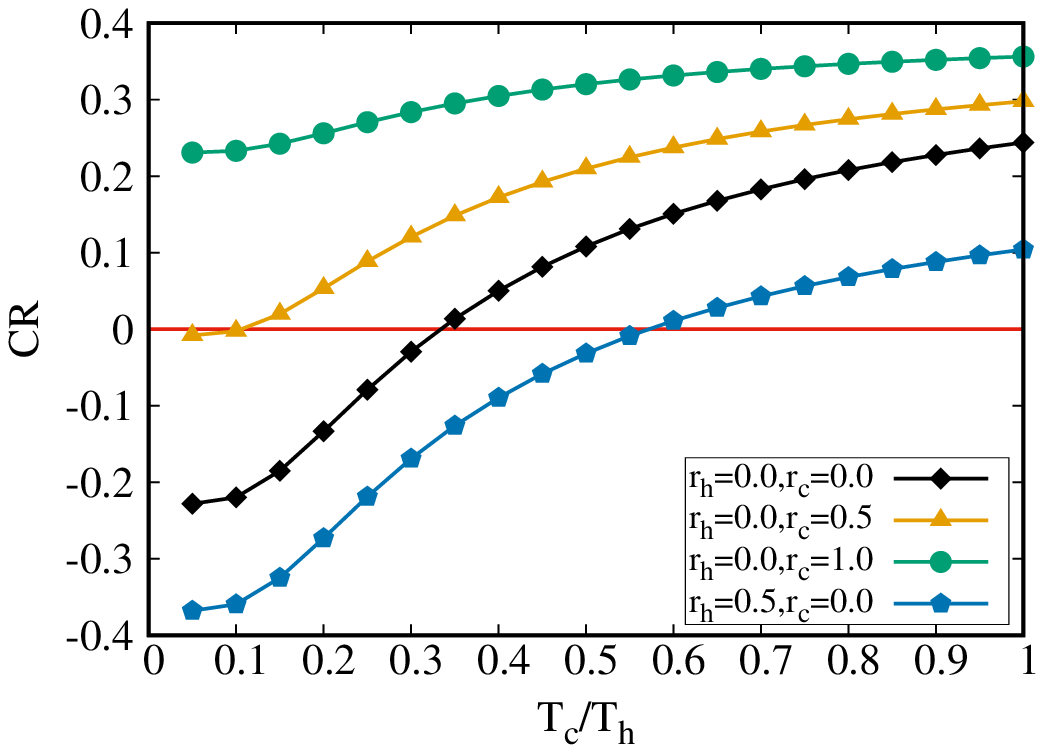}
	\caption{}
\end{subfigure}
 \begin{subfigure}{0.48\textwidth}
     \centering
     \includegraphics[width=\textwidth]{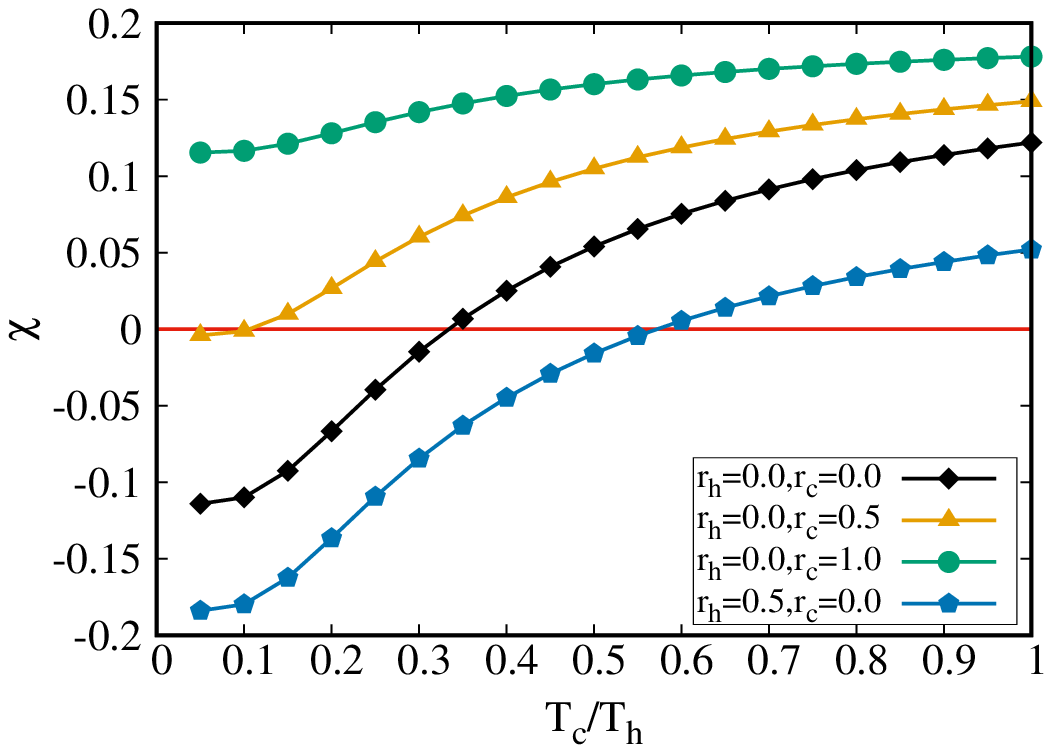}
     \caption{}
 \end{subfigure}
   \caption{ \label{fig:1q_cr_fig_tc_th_diff_r} Variation of (a) $CR$ and (b) $\chi$ of an OQR as a function of the temperature ratio $T_c/T_h$ for different sets of $r_h$ and $r_c$. Other parameters are : $\omega_h=30$, $\omega_c=10$, $\gamma_0=1$.}
 
  \end{figure}

\paragraph{Cooling rate of Refrigerator:-} Fig. \ref{fig:1q_cr_fig_tc_th_diff_r} (a) shows the dependence of $CR$ on the temperature ratio  $T_c/T_h$, for different values of the squeezing parameters $r_c$ and $r_h$. When the temperature ratio increases, i.e., the temperature  difference reduces, the $CR$ is observed to increase. This is in accordance with our expectations: a decrease in thermal gradient helps the refrigerator to pump heat against it. The curves shown in the figure are for different values of the squeezing parameters $r_h$ and $r_c$, as mentioned in the legends. 
The device no longer functions as a refrigerator below the line $CR=0$ (horizontal red line). The third curve from the top (black line with solid squares) is that of a thermal quantum refrigerator in the absence of squeezing, which serves as a reference. 
As long as $r_h<r_c$, the refrigerator performs better than the thermal refrigerator, implying that the effective temperature of the cold bath has increased with respect to the increase in the temperature of the hot bath. 
In the opposite regime, $r_h>r_c$ (cyan curve with pentagons), the refrigerator becomes more inefficient than the normal refrigerator. 
Fig. \ref{fig:1q_cr_fig_tc_th_diff_r} (b) shows the variation of the figure of merit $\chi$ with $T_c/T_h$ for the same set of parameters. Note that with $\omega_c/\omega_h$ fixed, the coefficient of performance $\zeta$ becomes a constant. Thus, from Eq. \eqref{eq:Ana_cr_1q} and \eqref{eq:chi}, it is easy to see that $\zeta$ will be proportional to $\chi$, thereby exhibiting exactly the same qualitative trends.

\begin{figure}[!ht]
 \centering
 	\begin{subfigure}{0.48\textwidth}
	\includegraphics[width=\textwidth]{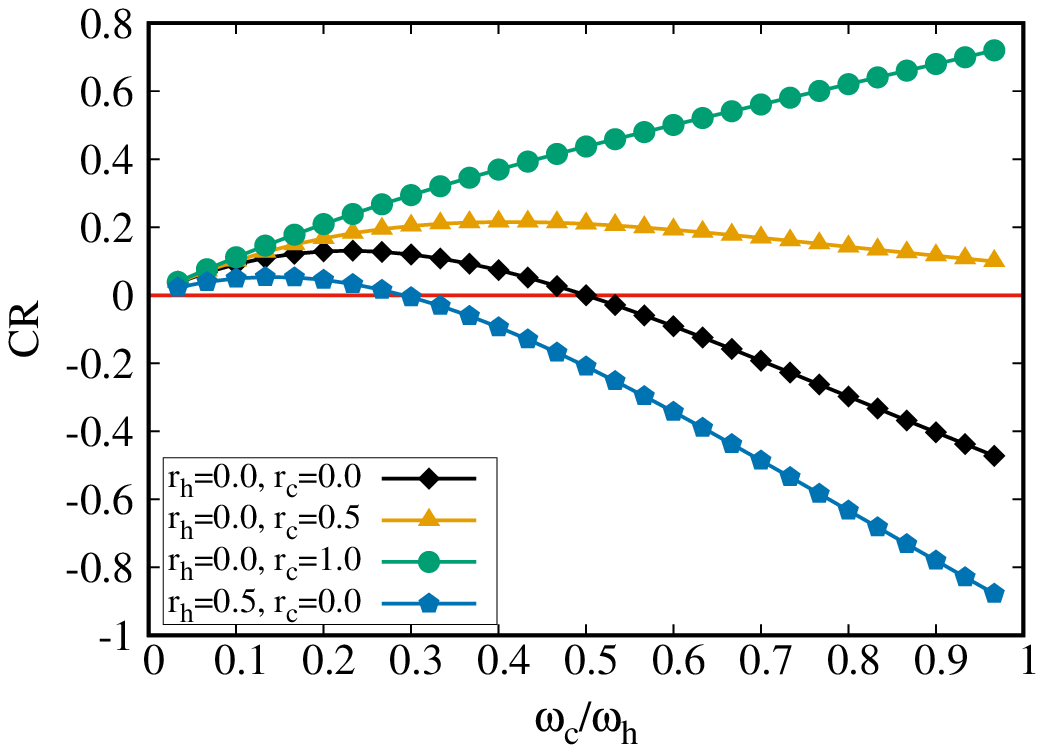}
	\caption{}
\end{subfigure}
 \begin{subfigure}{0.48\textwidth}
     \centering
     \includegraphics[width=\textwidth]{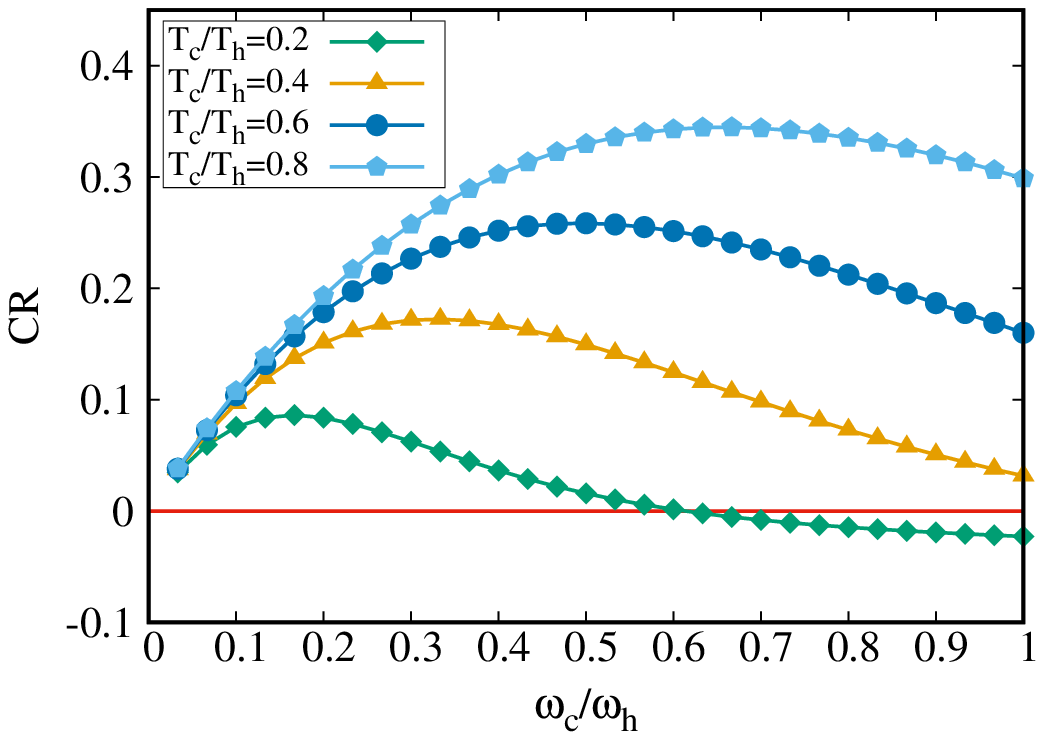}
     \caption{}
 \end{subfigure}
    \caption{(a) Variation of $CR$ of an OQR as a function of the $\omega_c/\omega_h$ for different $r_c$ and $r_h$. Other parameters are $T_h=20$, $T_c=10$, $\gamma_0=1$.
  (b) Variation of the $CR$ as the function of $\omega_c/\omega_h$ for different $T_c/T_h$ at  $r_c=0.5$ and $r_h=0$. }
   \label{fig:1q_cr_ratio_freq_temp}
  \end{figure}

In figure \ref{fig:1q_cr_ratio_freq_temp}(a), variation of $CR$ as a function of the ratio of energy gaps $\omega_c$/$\omega_h$ has been shown for different sets of the squeezing parameter $r_c$ and $r_h$. The other parameters are as mentioned in the figure caption. It can be clearly seen that whether the curve shows non-monotonicity or acts as a refrigerator at all, strongly depends on the ratio of energy gaps. As $r_c$ increases compared to $r_h$, the range of values of $\omega_c/\omega_h$ in which the system is in the refrigerator mode (i.e., the curves remain above the $CR=0$ line) increases. For high enough values of $r_c$, the non-monotonicity vanishes completely. The black curve, as before, is the reference line in the absence of squeezing: $r_h=r_c=0$. As expected, it is observed that the case in which $r_h$ exceeds $r_c$ under-performs with respect to this normal refrigerator, due to the increased thermal drive acting on the system in the opposite direction. 
Fig. \ref{fig:1q_cr_ratio_freq_temp}(b) shows the functional dependence of $CR$ again on the ratio $\omega_c/\omega_h$, but this time the squeezing parameters are kept constant and the temperature ratio $T_c/T_h$ is varied to obtain the different curves. The non-monotonicity in all the curves is apparent. 
These plots can be used to obtain the figure of merit at the maximum cooling rate, $\chi_{\rm MCR}$, as well as the coefficient of performance at the maximum cooling rate, $\zeta_{\rm MCR}$. These are useful parameters that quantify the performance of a refrigerator \cite{yan1990class,velasco1997new,allahverdyan2010optimal,de2012optimal}, similar to the efficiency at maximum power for an engine. These parameters have been explored next.

\begin{figure}[!ht]
 \centering
\begin{subfigure}{0.48\textwidth}
     \centering
     \includegraphics[width=\textwidth]{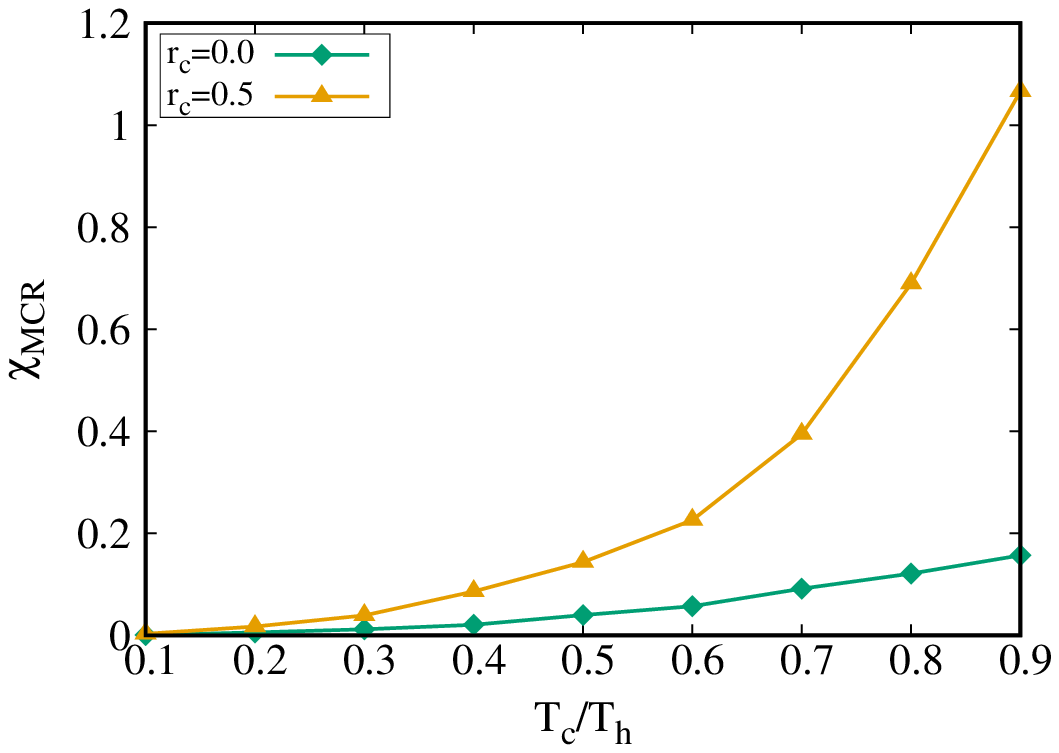}
     \caption{}
 \end{subfigure}
 \begin{subfigure}{0.48\textwidth}
     \centering
     \includegraphics[width=\textwidth]{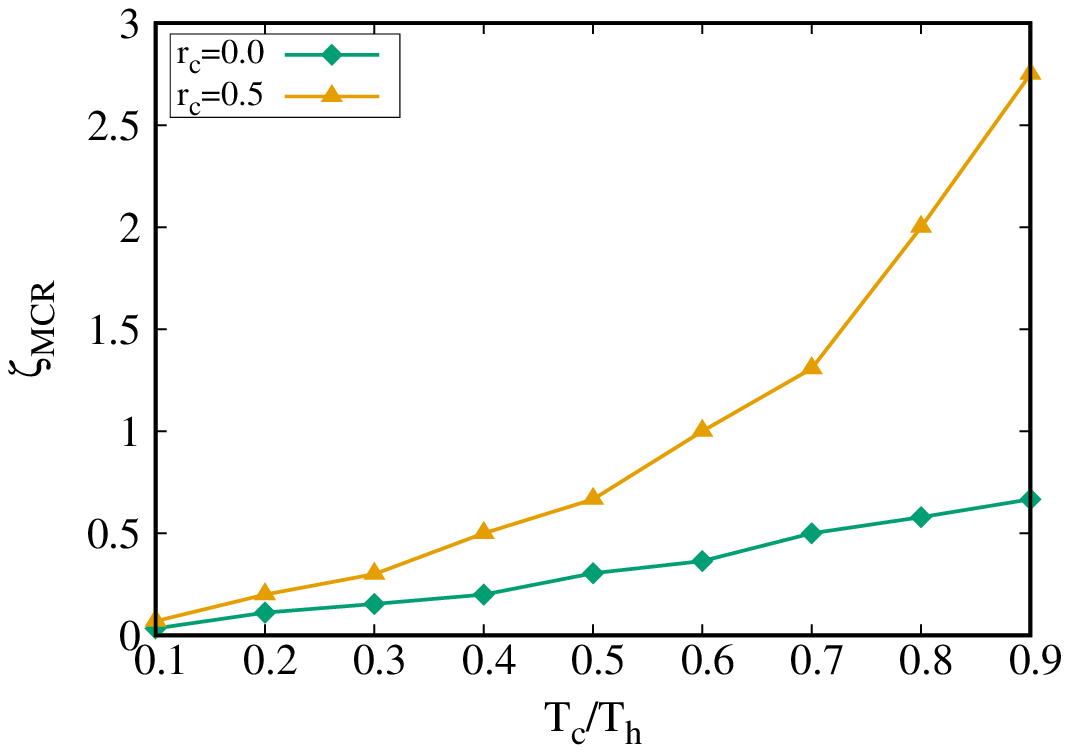}
     \caption{}
 \end{subfigure}
\caption{(a) Variation of $\chi_{MCR}$ with $T_c/T_h$, and (b) Variation of $\zeta_{\rm MCR}$ with $T_c/T_h$, for $r_c=0$ and 0.5. Other parameters are  $r_h=0$, $\gamma_0=1$,  $\omega_{c}=10$, $\omega_{h}=30$.}
 \label{fig:1q_cop_fig.m_CRmax}
 \end{figure}

\paragraph{$\chi$ and $\zeta$ at the maximum cooling rate:}  
Figs. \ref{fig:1q_cop_fig.m_CRmax}(a) and (b) show the variations in $\chi_{\rm MCR}$ and $\zeta_{\rm MCR}$ respectively with the ratio of bath temperatures. Both the quantities increase with the increase in $T_c/T_h$, and the increase is higher when the squeezing of the cold bath is more (i.e., the value of $r_c$ is higher). This is because, as can be observed from the plots of Fig. \ref{fig:1q_cr_ratio_freq_temp}(b), the maximum $CR$ occurs at values of $\omega_c/\omega_h$ that increase with the increase in $T_c/T_h$. Since $\zeta$ monotonically increases with $\omega_c/\omega_h$ (see Fig. \ref{fig:benchmarking}(a)), it can be readily inferred that the values of $\zeta$ at these maxima must increase as well with $T_c/T_h$.

 \begin{figure}[!ht]
 \centering
 	 \begin{subfigure}{0.48\textwidth}
     \centering
     \includegraphics[width=\textwidth]{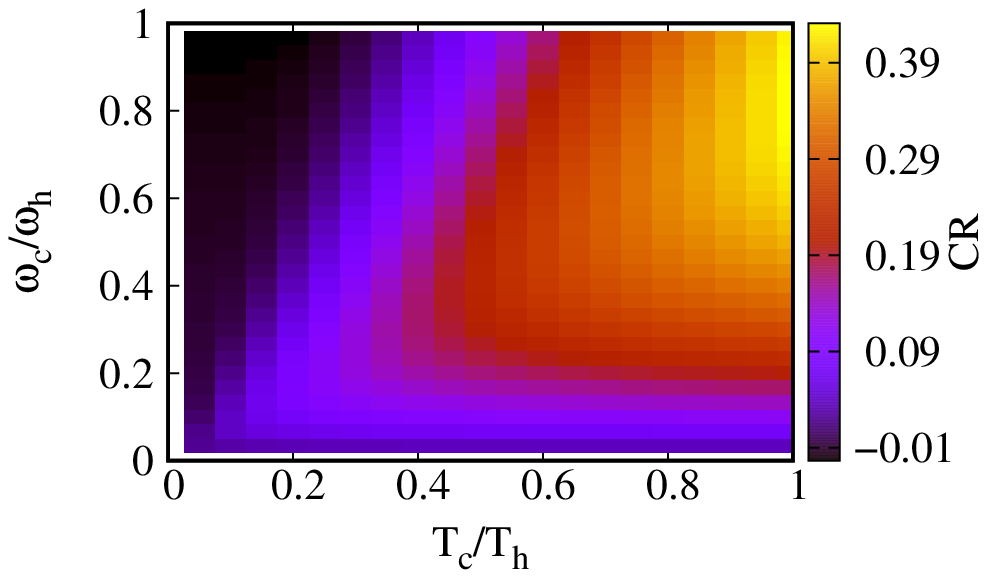}
     \caption{}
 \end{subfigure}
 \begin{subfigure}{0.5\textwidth}
     \centering
     \includegraphics[width=\textwidth]{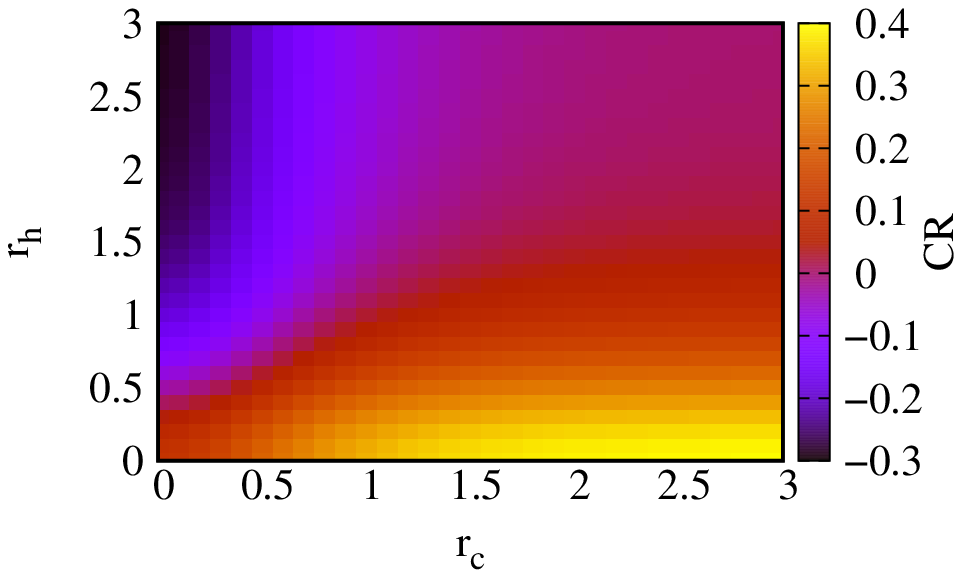}
     \caption{}
 \end{subfigure}
 \caption{(a) Phase plot of $CR$ of an OQR as a function of $T_c/T_h$ and $\omega_c/\omega_h$. The parameters used are $r_c=0.5$, and $r_h=0$. (b) Phase plot showing the $CR$ varies depending on the $r_c$ and $r_h$. Other parameter are $\gamma_0=1$, $\omega_c=10$, $\omega_h=30$, $T_h=20$, $T_c=10$.}
 \label{fig:ph_1}
 \end{figure}

\paragraph{Phase plots:} 
 The phase plot of OQR, showing the variation of $CR$ with the values of $\omega_c/\omega_h$ (vertical axis) and $T_c/T_h$ (horizontal axis), has been provided in Fig. \ref{fig:ph_1}(a). The non-monotonic behavior with respect to $\omega_c/\omega_h$ for fixed values of $T_c/T_h$ is apparent, which is consistent with the results shown in Fig. \ref{fig:1q_cr_ratio_freq_temp}(b).  Fig. \ref{fig:ph_1}(b) gives the phase plot of the variations in $CR$ with the squeezing parameters $r_h$ and $r_c$. Agreement with Fig. \ref{fig:1q_cr_ratio_freq_temp}(a) is observed in this case, showing that the effective temperature of a bath rises with the increase in squeezing, and vice versa.

\section{Results and Discussions on TQR} \label{sec:TQR}

We now examine the two-qubit refrigerator, whose dynamics were explained in Sec. \ref{subsec:two-qubit}, and compare the thermodynamic observables with those of OQR. We demonstrate that the TQR turns out to be a better choice in terms of the usefulness of the refrigerator. The parameter $\phi$ is found to have no effects on the refrigerator outputs, just as in the case of the OQR. Consequently, we again deal only with the parameter $r$, with subscripts $h$ and $c$ indicating the concerned bath. The other parameters that we keep fixed are the time duration of each stroke of the refrigerator cycle, $\tau=2$, and the spontaneous emission rates $\Gamma_1=\Gamma_2=\Gamma$, which we set to unity (refer to Eq. \eqref{eq:Gamma}). 
The working of a TQR can be divided into two regimes.
If the normalized distance $r_{12}$ (given by the ratio of the actual distance to $2\pi/k_0$) between the qubits is large, $r_{12}>1$, the regime is called \emph{independent} decoherence regime, while if $r_{12}\ll 1$, the regime is called \emph{collective} decoherence regime.

\begin{figure}[!ht]
	\begin{center}
	\includegraphics[width=0.5\textwidth]{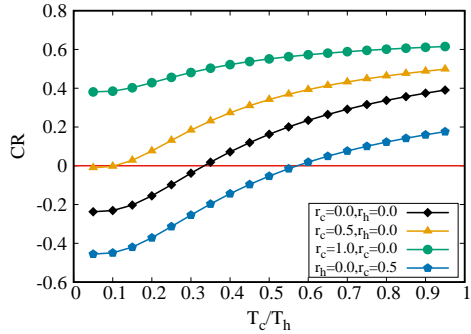}
\caption{\label{fig:2q_temp_diff_sq}Variation of $CR$ of a TQR as a function of the temperature ratio $T_c/T_h$ for different squeezed parameters $r_h$ and $r_c$. Other parameters are : $\omega_{0h}=30$, $\omega_{0c}=10$, $\Gamma=1$, $r_{12}=0.5$. }
	\setlength{\belowcaptionskip}{1pt}
	\end{center}
 \end{figure}

\paragraph{Cooling rate :-} Fig. \ref{fig:2q_temp_diff_sq} shows the variation of $CR$ with the ratio of bath temperatures $T_c/T_h$ in the decoherence regime. We find the trend to be very similar to that of Fig. \ref{fig:1q_cr_ratio_freq_temp}. However, there are substantial quantitative differences in favor of the TQR, as can be observed from the high values of $CR$ reached for similar values of the other parameters (see caption for the values of the set of parameters used). The effect of different combinations of $r_h$ and $r_c$ once again underscores the role played by these parameters in deciding the effective temperatures of the heat baths.

\begin{figure}[t]
 \centering
 	\begin{subfigure}{0.48\textwidth}
	\includegraphics[width=\textwidth]{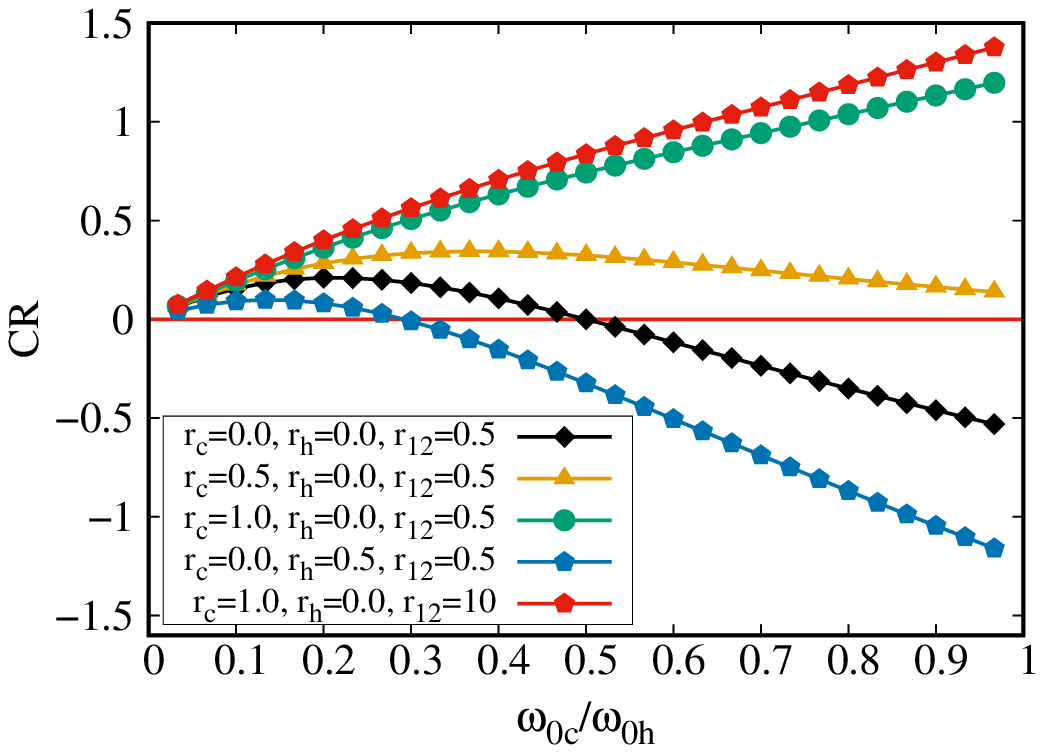}
	\caption{}
\end{subfigure}
 \begin{subfigure}{0.48\textwidth}
     \centering
     \includegraphics[width=\textwidth]{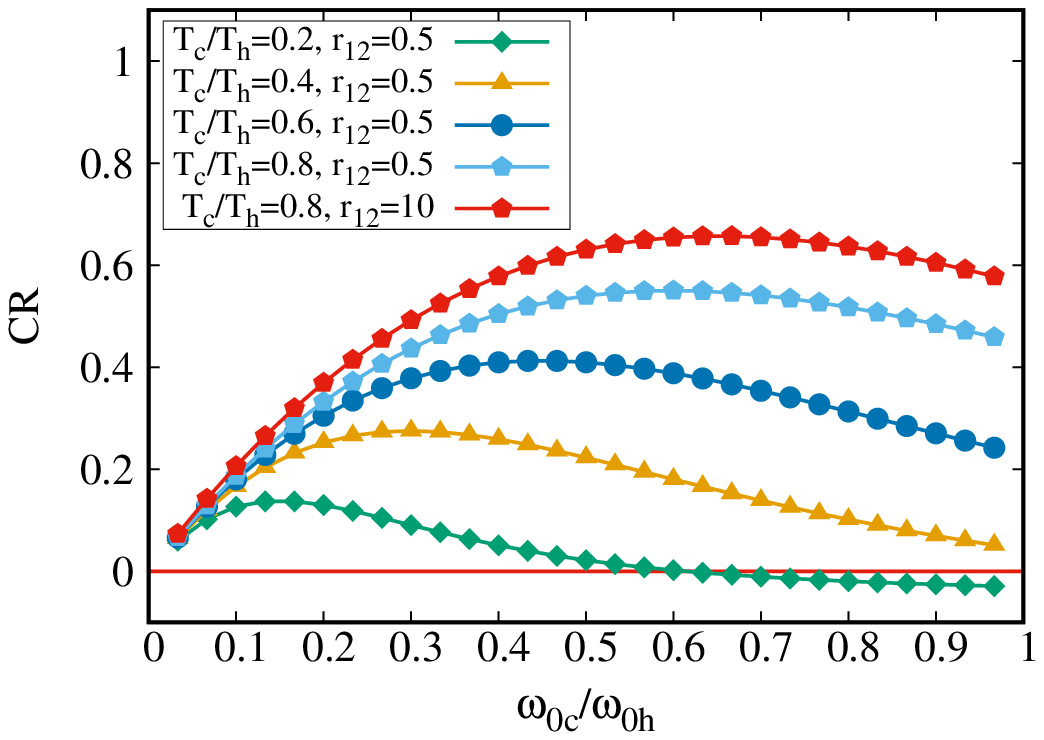}
     \caption{}
 \end{subfigure}
   \caption{(a)Variation of $CR$  of a TQR as a function of $\omega_c/\omega_h$ for different values of $r_c$ and $r_h$.
  (b) $CR$ as the function of $\omega_c/\omega_h$ for different values of $T_c/T_h$ with $r_c=0.5$ and $r_h=0$. Other parameters are $T_h=20$, $T_c=10$, $\Gamma=1$, $r_{12}=0.5$ for four curve from bottom and $r_{12}=10$ for first curve from top.}
   \label{fig:2q_freq_sq}
  \end{figure}

 Figs. \ref{fig:2q_freq_sq}(a) and (b) show the variations in $CR$ with the ratios $\omega_{0c}/\omega_{0h}$ and $T_c/T_h$ respectively, for the same parameters used for OQR. In both the subfigures, the first four curves from below are for the qubit being in the collective decoherence regime ($r_{12}\ll 1$, whose value we have chosen to be 0.5) for fixed values of squeezing parameters, while the upper plot (red solid line with solid pentagons) shows this functional dependence in the independent decoherence regime ($r_{12}>1$, whose value we have chosen to be 10). In the independent decoherence regime, the TQR is found to yield higher values of $CR$. This effect can be explained as follows. As shown in \cite{ficek2002entangled}, the number of decay channels for the system is higher in the independent decoherence regime. This in turn implies that both the absorption and dissipation of heat take place at a higher rate in this regime as compared to the collective decoherence regime, thereby leading to enhanced values of $CR$.
 Comparing with Figs. \ref{fig:1q_cr_ratio_freq_temp}(a) and (b) respectively, similar qualitative trends but with higher values of $CR$ are observed. 
 
 The higher values of $CR$ for a TQR can be heuristically argued as follows. Since out of the total number of particles in the ensemble distributed in all four levels, the lower two levels contain a bigger fraction as compared to the fraction of particles present in the ground state of a two-level system \cite{lahiri2023thermodynamics,manzano2018squeezed}. Since the levels are closer when in contact with the cold reservoir, the above fact leads to a higher probability of particles getting excited by absorbing heat from the colder reservoir, thereby leading to higher values of $\langle Q_c\rangle$ for the TQR.

\begin{figure}[!ht]
 \centering
   \begin{subfigure}{0.48\textwidth}
     \centering
     \includegraphics[width=\textwidth]{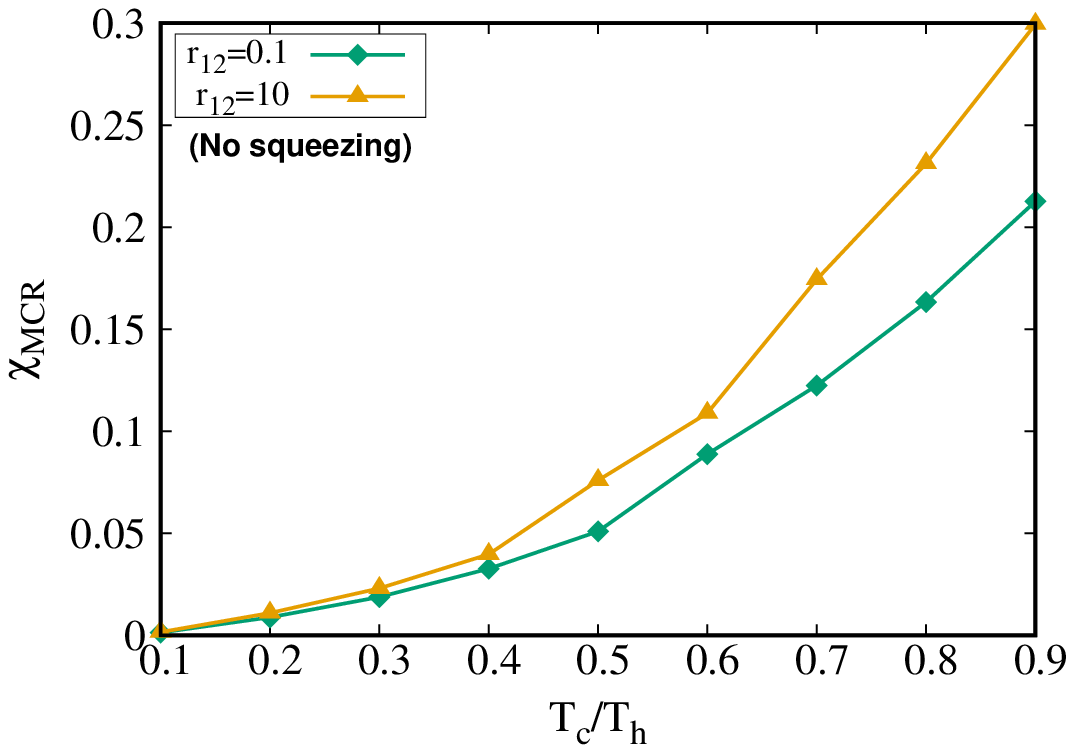 }
     \caption{}
 \end{subfigure}
  \begin{subfigure}{0.48\textwidth}
     \centering
     \includegraphics[width=\textwidth]{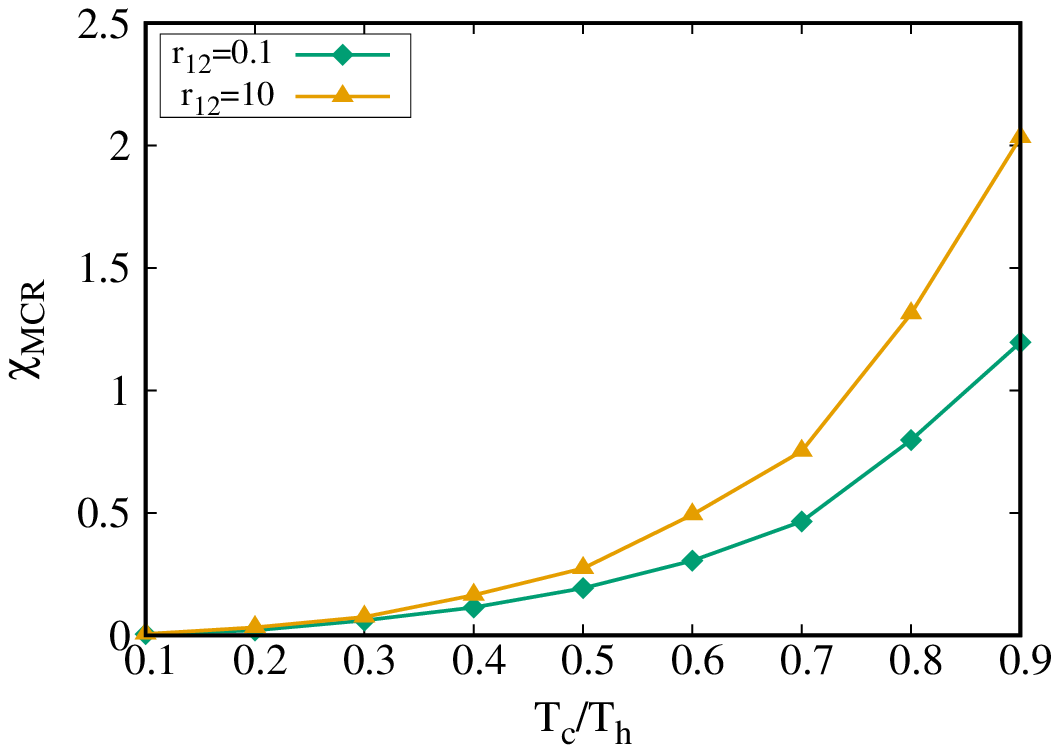}
     \caption{}
 \end{subfigure}
  \begin{subfigure}{0.48\textwidth}
     \centering
     \includegraphics[width=\textwidth]{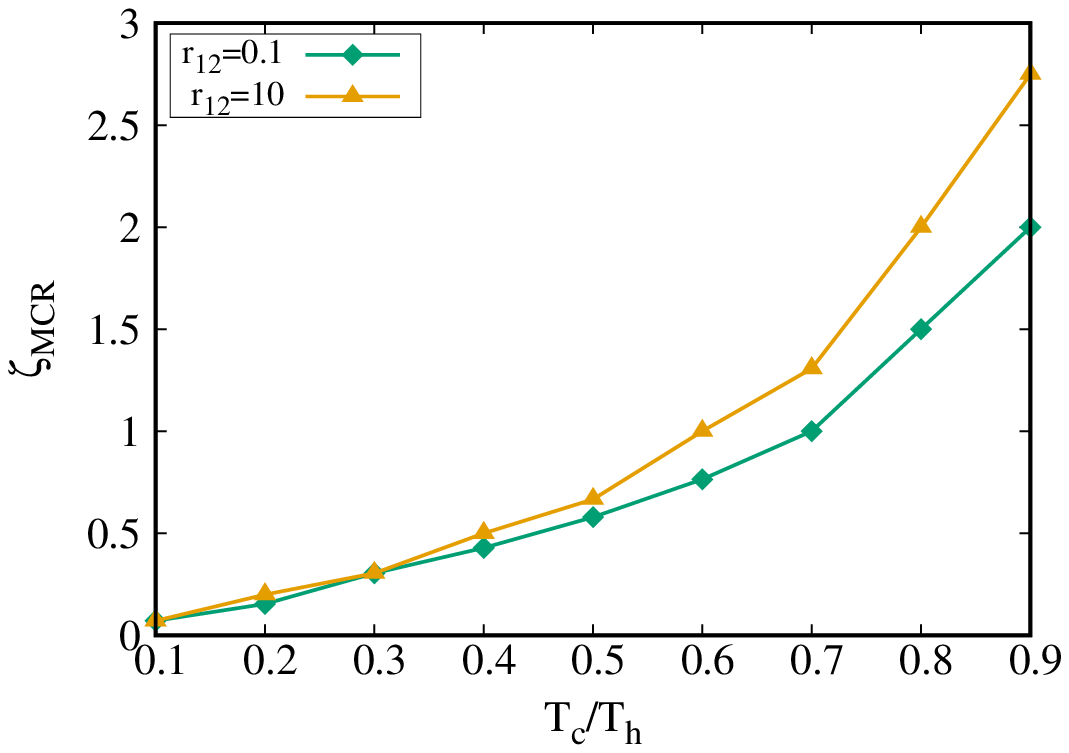}
     \caption{}
 \end{subfigure}
  \caption{ \label{fig:2q_fig.m_crmax}(a) Show the figure of merit $\chi_{MCR}$ as function of  $T_c/T_h$ in the collective regime ($r_{12}=0.1$) and independent decoherence regime ($r_{12}=10$) for  $r_h=0$ and $r_c=0$. (b) Same plot as in \ref{fig:2q_fig.m_crmax}(a) for  $r_h=0$ and $r_c=0.5$
  (c) Show the $\zeta_{MCR}$ as function of  $T_c/T_h$ in the collective decoherence regime ($r_{12}=0.1$) and independent decoherence regime ($r_{12}=10$) for $r_h=0$ and $r_c=0.5$. Other parameter are $\Gamma=1$, $\omega_{0c}=10$, $\omega_{0h}=30$.}
 \end{figure}

 \paragraph{$\zeta$ and $\chi$ at maximum cooling rate:} The procedure used to determine $\zeta_{\rm MCR}$   and $\chi_{\rm MCR}$ at the maximum cooling rate is the same as that used for the OQR. Fig. \ref{fig:2q_fig.m_crmax}(a) shows $\chi_{\rm MCR}$ as a function of $T_c/T_h$ for unsqueezed baths in the independent (upper curve) and collective (lower curve) decoherence regimes. Fig. \ref{fig:2q_fig.m_crmax}(b) provides similar curves when the cold bath is squeezed ($r_c=0.5$), showing substantial improvement over Fig. \ref{fig:2q_fig.m_crmax}(a). Fig. \ref{fig:2q_fig.m_crmax}(c) shows the variation of $\zeta_{\rm MCR}$ with the temperature ratio. In all the sub-figures, the independent decoherence regime is more conducive to the refrigerator's performance.
 Noting that since the maximum of $CR$ occurs at a higher value of $\omega_{0c}/\omega_{0h}$ in the independent decoherence regime (see Fig. \ref{fig:2q_freq_sq}(b)), we expect (one may refer to Eq. \eqref{eq:chi_max} for the quasistatically driven OQR as an approximate indicator of this qualitative nature) higher values of $\chi_{\rm MCR}$ and $\zeta_{\rm MCR}$ in the independent decoherence regime. The simulations  are observed to conform to our expectations.
 
\begin{figure}[!ht]
 \centering
 	\begin{subfigure}{0.49\textwidth}
	\includegraphics[width=\textwidth]{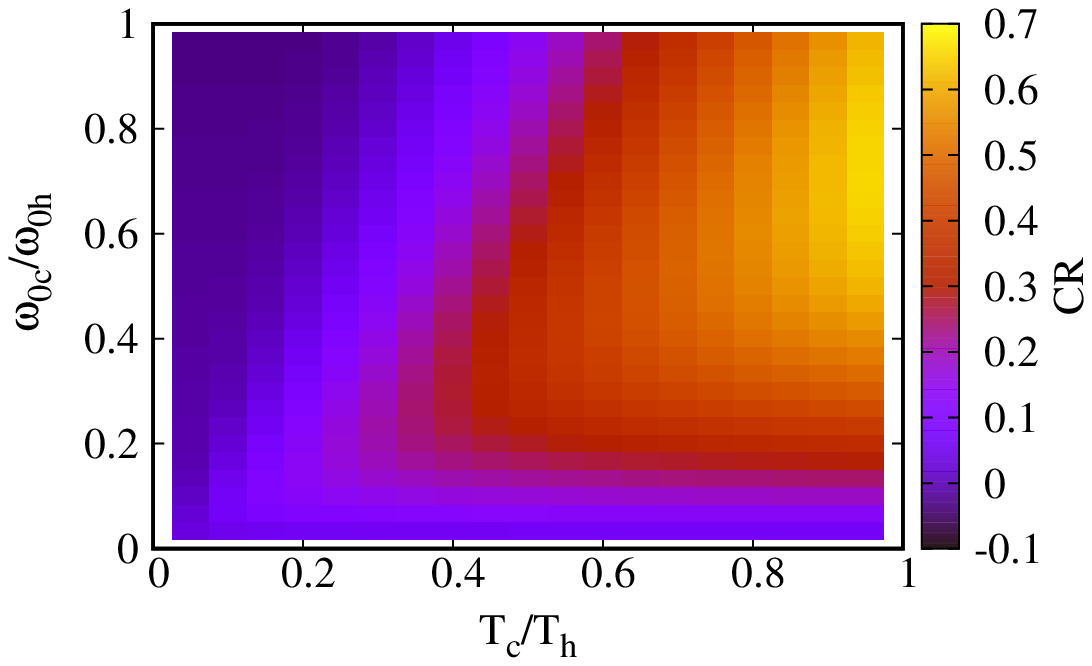}
	\caption{}
\end{subfigure}
 \begin{subfigure}{0.48\textwidth}
     \centering
     \includegraphics[width=\textwidth]{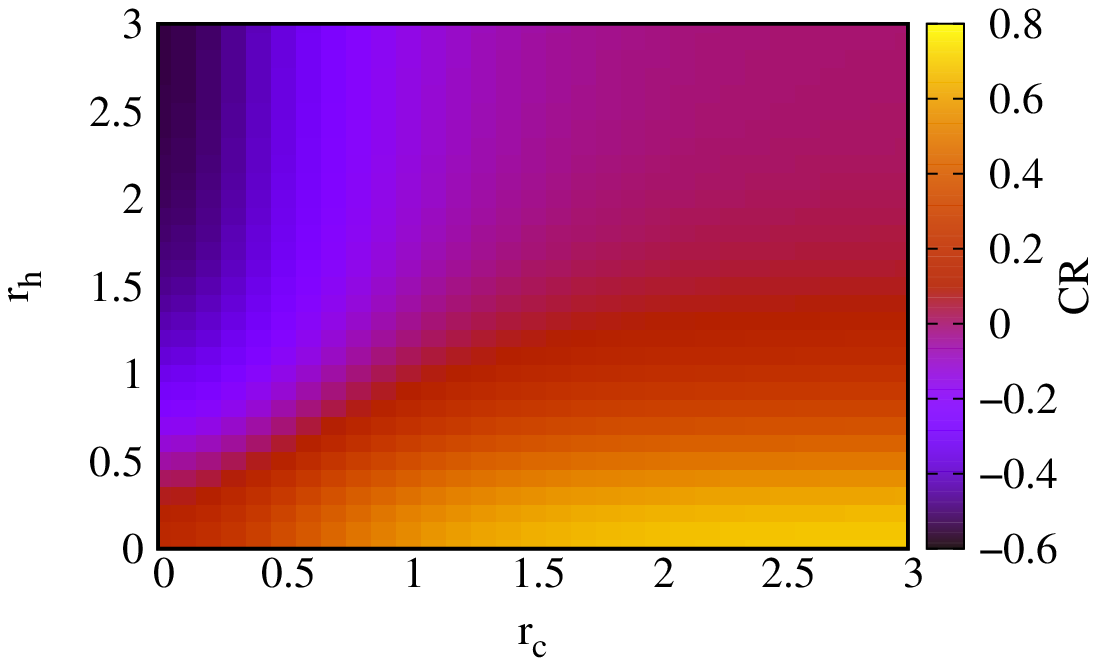}
     \caption{}
 \end{subfigure}
 \caption{(a) Phase plot of $CR$ of a TQR as a function of $T_c/T_h$ and $\omega_{0c}/\omega_{0h}$ for $r_c=0.5$, $r_h=0$. (b) Phase plot showing the $CR$ varies depending on the $r_c$ and $r_h$. Other parameter are $\Gamma=1$, $\omega_{0c}=10$, $\omega_{0h}=30$, $T_h=20$, $T_c=10$.}
\label{fig:ph_2}
\end{figure}

 In Fig. \ref{fig:ph_2}(a), the $CR$ as functions of $\omega_c/\omega_h$ and $T_c/T_h$ have been depicted by means of a phase plot. We observe that the non-monotonicity that was obtained in Fig. \ref{fig:ph_1}(a) for the OQR is also present in the case of the TQR. 
 Fig. \ref{fig:ph_2}(b) provides the functional dependence of $CR$ on $r_c$ and $r_h$ in the decoherence regime, which again shows similar qualitative trends as OQR. In both figures (a) and (b), the TQR quantitatively yields higher values of $CR$.

\section{Conclusions}\label{sec:conclusion}
 In this work, we have investigated one and two-qubit refrigerators in the presence of squeezed thermal reservoirs, and have made a comparative study of these systems. We have first described the model and theory of OQR and TQR. We have also derived the analytical expressions for the thermodynamics observables in the limit of quasistatic driving when squeezing is absent. Our simulations have been shown to agree with these results to very good accuracy.
 We have shown that the cooling rate $CR$ as well as the figure of merit $\chi$ increase monotonically with the temperature ratio $T_c/T_h$, for both OQR and TQR. 
 The coefficient of performance $\zeta$ is found to be dependent on the ratio of energy level spacings $\omega_c/\omega_h$ of the system, but  independent of the squeezing parameters. However, the squeezing parameters affect both $CR$ and $\chi$, the latter being proportional to the former, where the constant of proportionality is $\zeta$.
 As a function of the ratio of energy level spacings, the cooling rate shows a non-monotonic variation, with the general observation that it reaches higher values in the case of the TQR. These curves have been studied for different values of squeezing parameters as well as of $T_c/T_h$. From these plots, the values of $\chi_{\rm MCR}$ and $\zeta_{\rm MCR}$ have been extracted and plotted as a function of the temperature ratio. In accordance with our expectations, the curve for $r_c>0$ shows higher values of these parameters, for both OQR and TQR. 
 Phase plots have been plotted, showing the dependence of $CR$ on $T_c/T_h$ and $\omega_c/\omega_h$, as well as on $r_c$ and $r_h$. Once again, the TQR outperforms the OQR. An additional parameter in the case of TQR is the normalized distance $r_{12}$ between the two qubits. The values of $\chi_{\rm MCR}$ and $\zeta_{\rm MCR}$ are shown to be greater in the independent decoherence regime ($r_{12}>1$), as compared to the collective decoherence regime ($r_{12}\ll 1$), owing to the increased number of decay channels in the former case. It would be interesting to check whether increasing the number of spins of the working system induces its thermodynamic behavior to converge towards that of a multi-state system like the quantum harmonic oscillator \cite{wang2013efficiency,zhang2020optimization}.

\section{Acknowledgement}
One of us (AK)  thanks A. Kumari for providing valuable insights through stimulating discussions.
\appendix
\section{\label{app:cop} Derivation of coefficient of performance}

We carry out the derivation for the OQR in details, and then state the result for the TQR.
The average heat $\langle Q_{BC}\rangle$ absorbed from the cold bath in the system from B to C is given by
\begin{align}
 \av{Q_{BC}} &= Tr[(\rho_1-\rho_0)H(\omega_c)];
 \label{eq:Qc}
\end{align}
where $\rho_0=\rho_A=\rho_B$ and $\rho_1=\rho_C=\rho_D$ represent the density operators at A, B, C, and D.

Total work done is given by:
\begin{align}
\av{W} &= \av{W_{AB}}+\av{W_{CD}}\nonumber\\
       &=Tr[(\rho_1-\rho_0)\{H(\omega_{h})-H(\omega_c)\}]
       \label{eq:Work}
\end{align}
using the equation (\ref{eq:Qc}) and (\ref{eq:Work}), coefficient of performance $(\zeta)$ 
       \begin{align}
          \zeta &= \frac{\av{Q_c}}{\av{W}}\nonumber\\
               &=\frac{Tr[(\rho_1-\rho_0)H(\omega_c)]}{Tr[(\rho_1-\rho_0)\{H(\omega_{h})-H(\omega_c)\}]}\nonumber\\
               &=\frac{Tr[(\Delta\rho)H(\omega_c)]}{Tr[(\Delta\rho)\{H(\omega_{h})-H(\omega_c)\}]}
               \label{eq:COP}
       \end{align}
where, $(\rho_1-\rho_0)\equiv\Delta\rho$\\
  Now let
  \begin{align*}
      \Delta\rho &=
\begin{pmatrix}
  a & 0\\
 0 & b
 \end{pmatrix}.
 \end{align*}
 We also know,
 \begin{align*}
 H(\omega_c) &=
\begin{pmatrix}
 \omega_c/2 & 0\\
 0 & -\omega_c/2
\end{pmatrix},\nonumber\\
H(\omega_h) &=
\begin{pmatrix}
 \omega_h/2 & 0\\
 0 & -\omega_h/2
\end{pmatrix}.
\end{align*}
Then, 
\begin{align}
    \text{Tr}[(\Delta\rho)H(\omega_c)] =\omega_c\left(\frac{a-b}{2}\right)\nonumber\\
\text{Similarly}, \hspace{1cm} \text{Tr}[(\Delta\rho)H(\omega_h)] = \omega_h\left(\frac{a-b}{2}\right)
\end{align}
putting these relations into Eq. ($\ref{eq:COP}$), we get the coefficient of performance($\zeta$)
\begin{align}
    \zeta=\frac{\omega_c}{\omega_h-\omega_c}.
\end{align}
For the TQR, we have verified separately that $\zeta$ exhibits a similar functional dependence on $\omega_{0c}$ and $\omega_{0h}$:
\begin{align}
    \zeta \equiv \frac{\omega_{0c}}{\omega_{0h}-\omega_{0c}}.
    \label{eq:cop_two_equiv_one}
\end{align}


\begin{thebibliography}{10}

\bibitem{blickle2012realization}
Valentin Blickle and Clemens Bechinger.
\newblock Realization of a micrometer-sized stochastic heat engine.
\newblock {\em Nat. Phys.}, 8(2):143--146, 2012.

\bibitem{schmiedl2007efficiency}
Tim Schmiedl and Udo Seifert.
\newblock Efficiency at maximum power: An analytically solvable model for
  stochastic heat engines.
\newblock {\em EPL}, 81(2):20003, 2007.

\bibitem{kumari2020stochastic}
Aradhana Kumari, PS~Pal, Arnab Saha, and Sourabh Lahiri.
\newblock Stochastic heat engine using an active particle.
\newblock {\em Phys. Rev. E}, 101(3):032109, 2020.

\bibitem{kumari2021microscopic}
Aradhana Kumari and Sourabh Lahiri.
\newblock Microscopic thermal machines using run-and-tumble particles.
\newblock {\em Pramana}, 95:1--12, 2021.

\bibitem{long2015performance}
Rui Long and Wei Liu.
\newblock Performance of quantum otto refrigerators with squeezing.
\newblock {\em Phys. Rev. E}, 91(6):062137, 2015.

\bibitem{bhat2014nanobots}
AS~Bhat.
\newblock Nanobots: the future of medicine.
\newblock {\em Int. J. Manag. Sci. Eng. Manag.}, 5(1):44--49, 2014.

\bibitem{manjunath2014promising}
Apoorva Manjunath and Vijay Kishore.
\newblock The promising future in medicine: nanorobots.
\newblock {\em j. biomed. sci. eng.}, 2(2):42--47, 2014.

\bibitem{freitas2006pharmacytes}
Robert~A Freitas.
\newblock Pharmacytes: An ideal vehicle for targeted drug delivery.
\newblock {\em J. Nanosci. Nanotechnol.}, 6(9-10):2769--2775, 2006.

\bibitem{liang2000thermal}
Shoudan Liang, David Medich, Daniel~M Czajkowsky, Sitong Sheng, Jian-Yang Yuan,
  and Zhifeng Shao.
\newblock Thermal noise reduction of mechanical oscillators by actively
  controlled external dissipative forces.
\newblock {\em Ultramicroscopy}, 84(1-2):119--125, 2000.

\bibitem{briegel2008entanglement}
Hans~J Briegel and Sandu Popescu.
\newblock Entanglement and intra-molecular cooling in biological systems? {A}
  quantum thermodynamic perspective.
\newblock {\em arXiv:0806.4552}, 2008.

\bibitem{rana2014single}
Shubhashis Rana, PS~Pal, Arnab Saha, and AM~Jayannavar.
\newblock Single-particle stochastic heat engine.
\newblock {\em Phys. Rev. E}, 90(4):042146, 2014.

\bibitem{linden2010small}
Noah Linden, Sandu Popescu, and Paul Skrzypczyk.
\newblock How small can thermal machines be? {T}he smallest possible
  refrigerator.
\newblock {\em Phys. Rev. Lett.}, 105(13):130401, 2010.

\bibitem{brunner2014entanglement}
Nicolas Brunner, Marcus Huber, Noah Linden, Sandu Popescu, Ralph Silva, and
  Paul Skrzypczyk.
\newblock Entanglement enhances cooling in microscopic quantum refrigerators.
\newblock {\em Phys. Rev. E}, 89(3):032115, 2014.

\bibitem{rossnagel2014nanoscale}
Johannes Ro{\ss}nagel, Obinna Abah, Ferdinand Schmidt-Kaler, Kilian Singer, and
  Eric Lutz.
\newblock Nanoscale heat engine beyond the carnot limit.
\newblock {\em Phys. Rev. Lett.}, 112(3):030602, 2014.

\bibitem{abah2016optimal}
Obinna Abah and Eric Lutz.
\newblock Optimal performance of a quantum otto refrigerator.
\newblock {\em EPL}, 113(6):60002, 2016.

\bibitem{scovil1959three}
Henry~ED Scovil and Erich~O Schulz-DuBois.
\newblock Three-level masers as heat engines.
\newblock {\em Phys. Rev. Lett.}, 2(6):262, 1959.

\bibitem{alicki1979quantum}
Robert Alicki.
\newblock The quantum open system as a model of the heat engine.
\newblock {\em J. Phys. A}, 12(5):L103, 1979.

\bibitem{kosloff1984quantum}
Ronnie Kosloff.
\newblock A quantum mechanical open system as a model of a heat engine.
\newblock {\em chem. phys.}, 80(4):1625--1631, 1984.

\bibitem{zhang2020optimization}
Yanchao Zhang.
\newblock Optimization performance of quantum otto heat engines and
  refrigerators with squeezed thermal reservoirs.
\newblock {\em Physica A}, 559:125083, 2020.

\bibitem{wang2013efficiency}
Rui Wang, Jianhui Wang, Jizhou He, and Yongli Ma.
\newblock Efficiency at maximum power of a heat engine working with a two-level
  atomic system.
\newblock {\em Phys. Rev. E}, 87(4):042119, 2013.

\bibitem{vinjanampathy2016quantum}
Sai Vinjanampathy and Janet Anders.
\newblock Quantum thermodynamics.
\newblock {\em Contemp. Phys.}, 57(4):545--579, 2016.

\bibitem{lahiri2023thermodynamics}
Ashutosh Kumar, Trilochan Bagarti, Sourabh Lahiri, and Subhashish Banerjee.
\newblock Thermodynamics of one and two-qubit nonequilibrium heat engines
  running between squeezed thermal reservoirs.
\newblock {\em Physica A}, page 128832, 2023.

\bibitem{yan1990class}
Zijun Yan and Jincan Chen.
\newblock A class of irreversible carnot refrigeration cycles with a general
  heat transfer law.
\newblock {\em J. Phys. D: Appl. Phys}, 23(2):136, 1990.

\bibitem{singh2020optimal}
Varinder Singh, Tanmoy Pandit, and Ramandeep~S Johal.
\newblock Optimal performance of a three-level quantum refrigerator.
\newblock {\em Phys. Rev. E}, 101(6):062121, 2020.

\bibitem{ficek2002entangled}
Zbigniew Ficek and Ryszard Tana{\'s}.
\newblock Entangled states and collective nonclassical effects in two-atom
  systems.
\newblock {\em Phys. Rep}, 372(5):369--443, 2002.

\bibitem{banerjee2010dynamics}
Subhashish Banerjee, V~Ravishankar, and R~Srikanth.
\newblock Dynamics of entanglement in two-qubit open system interacting with a
  squeezed thermal bath via dissipative interaction.
\newblock {\em Ann. Phys.}, 325(4):816--834, 2010.

\bibitem{subhashish2019open}
Subhashish Banerjee.
\newblock {\em Open quantum system: Dynamics of nonclassical evolution}.
\newblock springer, 2019.

\bibitem{breuer2002theory}
Heinz-Peter Breuer and Francesco. Petruccione.
\newblock {\em The theory of open quantum systems}.
\newblock Oxford University Press, 2002.

\bibitem{manzano2018squeezed}
Gonzalo Manzano.
\newblock Squeezed thermal reservoir as a generalized equilibrium reservoir.
\newblock {\em Phys. Rev. E}, 98(4):042123, 2018.

\bibitem{velasco1997new}
Santiago Velasco, Jos{\'e}~MM Roco, Alejandro Medina, and A~Calvo
  Hern{\'a}ndez.
\newblock New performance bounds for a finite-time carnot refrigerator.
\newblock {\em Phys. Rev. Lett.}, 78(17):3241, 1997.

\bibitem{allahverdyan2010optimal}
Armen~E Allahverdyan, Karen Hovhannisyan, and Guenter Mahler.
\newblock Optimal refrigerator.
\newblock {\em Phys. Rev. E}, 81(5):051129, 2010.

\bibitem{de2012optimal}
C~De~Tom{\'a}s, A~Calvo Hern{\'a}ndez, and JMM Roco.
\newblock Optimal low symmetric dissipation carnot engines and refrigerators.
\newblock {\em Phys. Rev. E}, 85(1):010104, 2012.

\end{thebibliography}
\end{document}